# Achieving Data Utility-Privacy Tradeoff in Internet of Medical Things: A Machine Learning Approach

Zhitao Guan[1], Zefang Lv[2], Xiaojiang Du[3], Longfei Wu[4], Mohsen Guizani[5]

1. School of Control and Computer Engineering, North China Electric Power University, China
2. School of Mathematics and Physics, North China Electric Power University, China
3. Department of Computer and Information Science, Temple University, Philadelphia PA, USA
4. Department of Mathematics and Computer Science, Fayetteville State University, Fayetteville, NC, USA
5. Electrical and Computer Engineering Department, University of Idaho, Moscow ID, USA.

***Abstract*—The emergence and rapid development of the Internet of Medical Things (IoMT), an application of the Internet of Things into the medical and healthcare systems, have brought many changes and challenges to modern medical and healthcare systems. Particularly, machine learning technology can be used to process the data involved in IoMT for medical analysis and disease diagnosis. However, in this process, the disclosure of personal privacy information must receive considerable attentions especially for sensitive medical data. Cluster analysis is an important technique for medical analysis and disease diagnosis. To enable privacy-preserving cluster analysis in IoMT, this paper proposed an Efficient Differentially Private Data Clustering scheme (EDPDCS) based on MapReduce framework. In EDPDCS, we optimize the allocation of privacy budgets and the selection of initial centroids to improve the accuracy of differentially private K-means clustering algorithm. Specifically, the number of iterations of the K-means algorithm is set to a fixed value according to the total privacy budget and the minimal privacy budget of each iteration. In addition, an improved initial centroids selection method is proposed to increase the accuracy and efficiency of the clustering algorithm. Finally, we prove that the proposed EDPDCS can improve the accuracy of the differentially private k-means algorithm by comparing the Normalized Intra-Cluster Variance (NICV) produced by our algorithm on two datasets with two other algorithms.**

***Keywords—Differential privacy, K-Means clustering, Internet of Medical Things, Machine learning, MapReduce.***

## I. INTRODUCTION

In the past few years, the emergence of the Internet of Medical Things (IoMT) has attracted researchers from both the IT industry and the healthcare field [1]. There have been many IoMT-based platforms, applications, and services for remote health monitoring, fitness programs, chronic diseases, and elderly care [2] [3]. IoMT also provides alternative solutions to the problems faced by traditional medical systems, such as the lack of doctors, healthcare resources and research data.

Specifically, the rapid development of IoMT has improved the traditional medical systems in various aspects, such as disease diagnosis and analysis [4] [5]. The health data collected in IoMT can be used by researchers to diagnose and predict diseases. With the popularity of IoMT system as well as its terminal devices, the volume of data collected is also vastly growing. Therefore, the big data technologies have been applied to IoMT to process and analyze the health data, so that medical researchers can better conduct disease risk assessment and prediction [6] [7]. Cluster analysis is a typical unsupervised learning data mining method, which can be applied in disease diagnosis [8]. The main idea is to divide the data into several clusters so that the distances among data items of the same cluster are as small as possible while the distances among data items of different clusters are as large as possible. Medical researchers can obtain general distribution and clinical phenotypes of a disease by performing cluster analysis of medical big data [9][10]. Doctors or medical researchers can not only better diagnose diseases and treat patients through information obtained from clustering results, but also better study the causes of diseases and thus promote the development of medical services.

However, the health data is very sensitive when linked with individual users. If not properly handled, the collection and analysis of the health data may leak users' privacy information. Therefore, it is critical to achieve privacy-preserving data analysis in IoMT. Existing privacy-preserving methods on data analysis include *k*-anonymity [11], *l*-diversity [12], differential privacy [13], and so on. Cluster analysis also faces same challenge. Although the clustering results of health data can provide valuable information, it may also leak private information regarding a single record in the dataset, posing a threat to personal privacy. Moreover, privacy preserving cluster algorithms usually face the challenge to achieve the tradeoff between privacy and accuracy.

In addition, in this Big Data era, the size of the dataset is getting larger and larger. As a consequence, the computing capability of a single computer is difficult to meet the needs. Therefore, developing privacy-preserving cluster analysis techniques and increasing the computational speed of cluster analysis are the two important and urgent tasks in IoMT [14].

In order to address the privacy preservation as well as the decreasing accuracy caused by privacy preserving techniques and the computational efficiency issues of cluster analysis in IoMT, we proposed an Efficient and Differentially Private Data Clustering scheme (EDPDCS) with high accuracy and efficiency while preserving privacy of data contributors. The main contributions of our paper are as follows:

1. We proposed an efficient privacy-preserving data clustering algorithm EDPDCS over the MapReduce framework for IoMT. The number of iterations of the K-means algorithm is set to a fixed value according to the total privacy budget and the minimal privacy budget allocated to each iteration calculated by analyzing the mean squared error (MSE) between noisy centroids and true centroids in one iteration.
2. We deployed an improved initial centroids selection algorithm in the MapReduce framework by selecting a small portion of the dataset and performing rough clustering in advance to select the initial centroids to improve the accuracy of differentially private K-means algorithm. And we developed a method for selecting the initial centroid for a specified number of clusters $k$ to solve the problem that the number of points outputs by the canopy algorithm is uncertain.
3. By experimenting with two datasets containing personal information in the medical and health filed, we verified that our proposed algorithm EDPDCS can reduce the Normalized Intra-Cluster Variance (NICV) and improve the accuracy of the clustering results while preserving personal privacy.

The rest of our paper is arranged in the following order. In section 2, we discuss the related work. Our model and design goals are introduced in section 3. In section 4, we introduce the differential privacy and two major algorithms, including the K-means algorithm and canopy algorithm. Our proposed efficient and differentially private K-means algorithm based on MapReduce is described in section 5. Section 6 presents the privacy analysis of our proposed algorithm. In section 7, we evaluate the accuracy and efficiency of the proposed algorithm. In section 8, we conclude the paper.

## II. Related Work

The emergence and rapid development of IoMT has brought many changes to the medical and health field. It has attracted the attention of many researchers.

There have been many research work on IoMT-based platforms, applications, and services, such as remote health monitoring, fitness programs, chronic diseases, and elderly care [15] [16] [17] [18] [19]. In [15], a novel joint IoMT and Product Lifecycle Management (PLM) based framework is proposed for medical healthcare applications to regulate the information transfer from one entity to another and between devices in an efficient and accurate way, while managing the battery lifecycle and energy of the resource-constrained tiny wearable devices.

The health data of users involved in IoMT also has great research value. Big data technologies can be used to analyze health data to assist medical personnel in disease diagnosis and analysis. There have been many studies in this area [7] [9] [20] [21]. Reference [7] studied the analysis and management issue of the large-scale data in the health field, the main purpose of their proposal is to first collect the medical (e-health) big data in real time, then process and analyze the data in the cloud. Among big data technologies, cluster analysis can be used to analyze not only the general distribution of a disease in case of some factors, such as gender and age, but also the clinical phenotypes. Reference [9] studied clinical phenotypes of Nasal Polyps and Comorbid Asthma based on cluster analysis of disease history.

However, sensitive health data relevant to users' privacy information may be involved when applying big data technologies to analyze the collected users' health data. There are some existing solutions to the security and privacy issues in IoMT [22] [23] [24] [25] [26]. Reference [26] introduced a clustering-based K-anonymity method as the building block of privacy preserving for data collected by medical wearable devices.

In the research on data privacy preservation, traditional methods, like $k$-anonymity [11] and $l$-diversity [12], can only deal with attacks under specific background knowledge while practical attacks can be mounted against all these techniques [27] [28]. The differential privacy technique proposed in [10] can preserve privacy for all individual contributors in a dataset. Differential privacy is a privacy protection method in which random noise following a specific distribution is added to distort the data [29]. It has been increasingly adopted in data analysis to preserve individual privacy [30] [31].

Various privacy-preserving K-means clustering methods have been proposed [32] [33]. The clustering under differential privacy has also been studied. Two important issues of differentially private K-means are the allocation of privacy budget and the initial centroids selection. There are generally two different methods for the allocation of privacy budgets in each iteration of the clustering algorithm, which correspond to two ways to determine the number of iterations including fixed iterations and unfixed iterations. [34] proposed an improved K-means clustering algorithm which satisfies differential privacy. The authors developed techniques to analyze MSE between the noisy centroids and the true centroids in one iteration and used this technique to determine the number of iterations and the budget allocation. As for the initial centroid selection, there are some methods including random selection, dividing dataset into subsets of equal size and finding the center points. [35] proposed a DPLK-means algorithm based on differential privacy, which improved the selection of the initial center points through performing the differential privacy K-means algorithm to each subset divided by the original dataset.

To increase the computational speed of cluster analysis, many studies have proposed to conduct cluster analysis on distributed computing platforms. Reference [36] developed a parallel K-means clustering algorithm based on MapReduce, which is a simple yet powerful parallel programming technique. In [36], Map function is used to perform the procedure of assigning each sample to the closest centroid while reduce function is used to update the new centroids. Several papers (e.g., [37-42]) have studied related issues.

## III. PRELIMINARIES

In this section, the differential privacy and two main algorithms are introduced which are used for differentially private K-means algorithm in MapReduce framework.

### A. Differential Privacy

Differential privacy protects individual privacy by adding noise to the query results, while maintaining the statistical characteristics and accuracy of the query results in an acceptable range.

**Definition 1** $\varepsilon$ -differential privacy：

A randomized mechanism M satisfies $\varepsilon$ -differential privacy if for any pair of neighboring datasets $D, D'$ differing in at most one data record and for any set of possible output $S \in Range(M)$ ,

$$\Pr(M(D) \in S) \le e^{\varepsilon} \cdot \Pr(M(D') \in S).$$

The privacy budget $\varepsilon$ represents the level of privacy guarantee - a lower privacy budget provides a stronger privacy guarantee.

### B. K-means Clustering Algorithm

Cluster analysis is a very important topic in data analysis. The purpose of clustering is to classify the data into different classes. The k-means clustering algorithm is the simplest and most commonly used clustering algorithm. The fundamental principle is to divide the data into k clusters on the basis of minimizing the error function, with distance as the rating index of similarity. That is, the shorter the distance is between two objects, the greater of similarity they have. Given a d-dimensional dataset $D = \{x^1, x^2, \cdots, x^N\}$ ($N$ is the total number of data points), the k-means algorithm divides the data points in D into k sets $\mathrm{O} = \{O^1, O^2, \cdots, O^k\}$ so that MSE within the cluster is minimized:

$$NICV = \frac{1}{N}\sum_{i=1}^{k}\sum_{x^l \in O^i}\left\|x^l - o^i\right\|^2, \ (1)$$

and

$$o_j^i = \frac{\sum_{x^l \in O^i} x_j^l}{\left|O^i\right|}, j = 1, 2, \cdots d.$$

### C. Canopy Clustering Algorithm

Canopy [43] is a fast, simple, but less accurate clustering algorithm. Canopy algorithm classifies a bunch of data into n data piles under certain rules, using two artificially determined thresholds $t_1$ and $t_2$ . Unlike traditional clustering algorithms, such as K-means, the biggest feature of the Canopy algorithm is that it does not need to specify the number of clusters $k$ in advance. Although it has lower precision, Canopy algorithm can speed up the clustering calculation. Therefore, it is usually used to perform a "rough" clustering of dataset to obtain $k$ and the set of centroids.

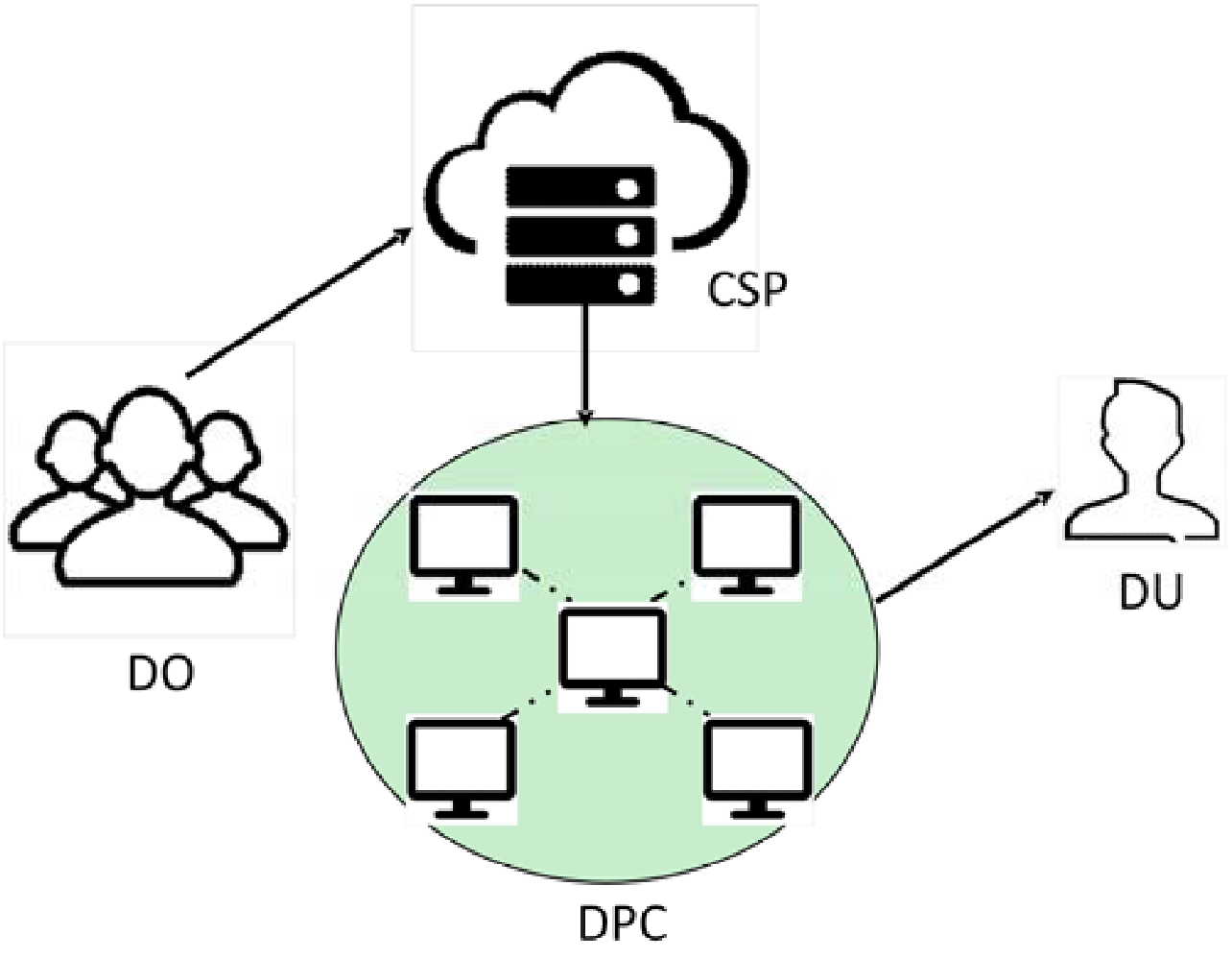

Fig. 1 Overview of data collection, transmission and processing

## IV. MODEL ANG DESIGN GOALS

### A. An Overview

IoMT is composed of a collection of terminal devices connected to the Internet, to provide medical and healthcare services. The analysis of health data for disease diagnosis and prediction is an important application in IoMT. As shown in Fig.1, four major entities are involved in Internet of Medical Things: Data Owner (DO), Cloud Service Provider (CSP), Data Processing Center (DPC) and Data User (DU).

1) DO: DO wears terminal devices equipped with health sensors, which will transport their health data to the cloud service provider. DO are assumed to be honest in the system.
2) CSP: The cloud stores all data uploaded by DO. CSP is assumed to be honest.
3) DPC: The data processing center processes the data and sends the results to the data users. DPC is assumed to be honest in the system.
4) DU: The data user receives the data processing results from the data processing center and performs tasks like disease diagnosis and prediction based on the results. DU is assumed untrusted.

However, the analysis process and the release of the analysis results may lead to the leakage of users' privacy information. In order to preserve privacy in the k-means clustering algorithm, we design a clustering algorithm satisfying the differential privacy. For computational efficiency, we deploy our algorithm in a distributed environment – MapReduce framework. And it is feasible to deploy a clustering algorithm that satisfies the differential privacy in the distributed environment. The differential privacy can be satisfied by adding Laplace noise to the centroid in Reduce task when calculating new centroids. Taking all these into considerations, we proposed an efficient differentially private data clustering scheme on the MapReduce framework to implement privacy-preserving cluster analysis in IoMT. We use MSE to determine the number of iterations. Additionally, we use an improved initial centroids selection algorithm based on canopy algorithm to initialize the centroids for K-means algorithm.

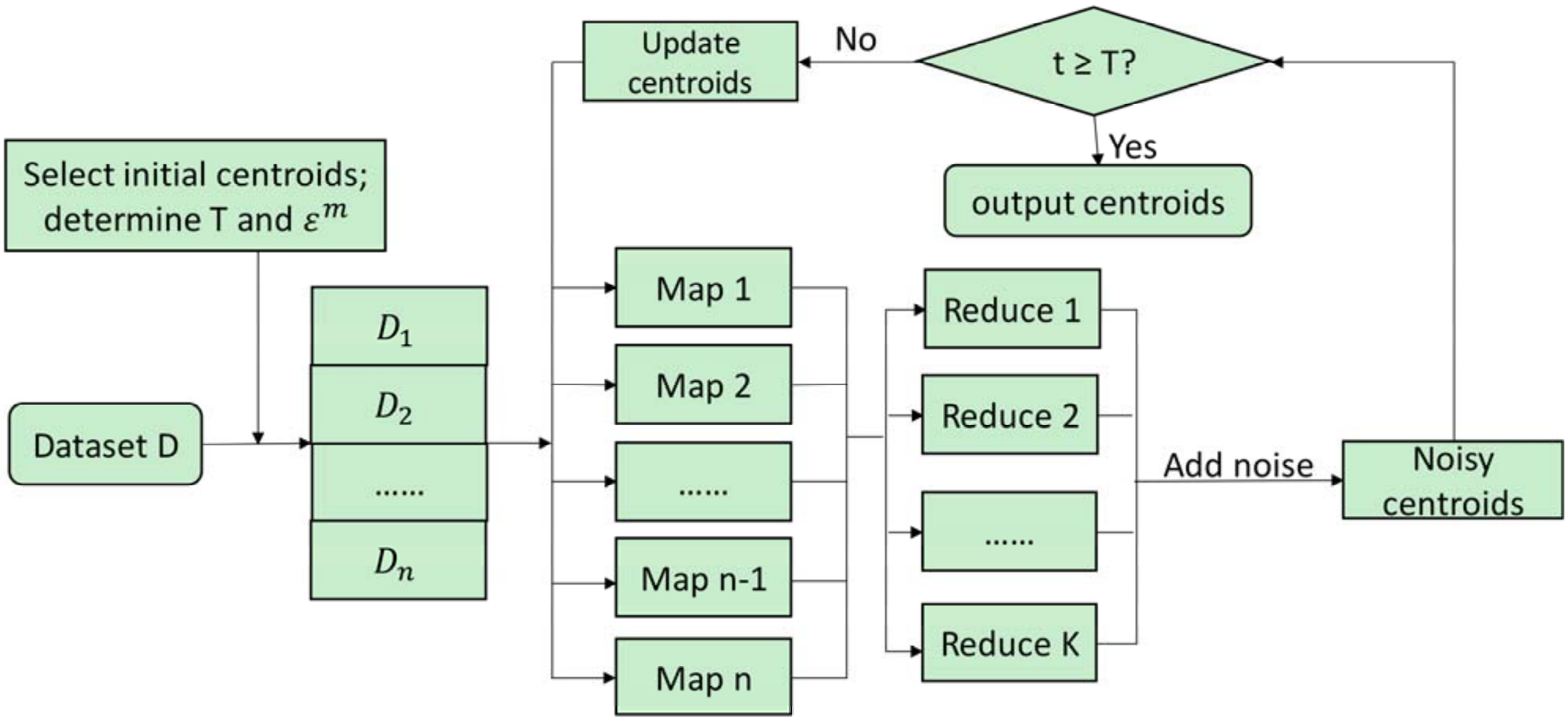

Fig. 2 Differentially Private K-means Algorithm in MapReduce Framework

### B. Design Goals

In order to solve the above issues, the design goal of our algorithm can be roughly divided into two aspects:

1) Privacy preservation: the change of cluster result in the centroids and the number of points in the clusters caused by the modification of the dataset will not reveal personal sensitive information. In other words, a malicious analyst cannot obtain any private information of a single record by mining a similar dataset, compared with original dataset.
2) Accuracy: achieve a tradeoff between accuracy of cluster results and privacy preservation.
3) Effectiveness: We conduct cluster analysis on a distributed computing platform-MapReduce framework to improve the computational speed for large-scale data.

### C. Security Model

In this subsection, we introduce the security model of our system. We assume that DO, CSP and DPC are trusted, and that DU is untrusted. In particular, DU may be a malicious analyst who try to explore sensitive information of DO by analyzing results from DPC. Differential privacy guarantees a strong privacy that deleting or adding a particular record in a dataset will not significantly change the output of any function on a dataset. Therefore, a malicious analyst or an adversary will just obtain approximate information about any individual record rather than specific information.

## V. Description of Our Scheme

The EDPDCS proposed in this paper is designed to ensure that the change of the centroid and the number of records of each cluster does not reveal private information when data is changed in the MapReduce distributed environment. A malicious analyst cannot obtain any private information of a single record by mining the data set.

The basic idea of the algorithm is to use the Mapper task on the distributed computing node to determine the cluster to which each record belongs. The Reducer task is employed to calculate the sum of the number of records in the cluster and the corresponding attributes. Then, Laplace noise is added to ensure that the results of the cluster analysis satisfy ε-differential privacy. Fig. 2 shows the whole idea of our proposed algorithm. Before performing the clustering algorithm, we calculate the minimum privacy budget that makes the algorithm satisfies differential privacy and achieve maximal accuracy of cluster result in case of total privacy budget and accuracy requirement, and then determine the number of iterations according to the minimum privacy budget. We developed an improved initial centroids selection algorithm derived from canopy algorithm to increase the accuracy of clustering results and developed a method for selecting the initial centroid for a specified number of clusters $k$ to solve the problem that the number of points outputs by the canopy algorithm is uncertain.

The proposed differentially private K-means algorithm mainly consists of three core parts. The first part is to deploy the K-means algorithm into the MapReduce framework. The second part is to add noise to the distributed k-means algorithm to satisfy the differential privacy and determine the condition of the algorithm iterations. Finally, the selection of the initial centroids will affect the results of the algorithm.

Therefore, how to select the initial centroids is also a part of the algorithm's considerations.

Let the total number of points in the data set be N, the number of iterations is T and each point is recorded as $x^i$ (1 ≤ i ≤ N), the dimension of each point is d, the centroids are recorded as $o^j$ (1 ≤j≤k) . Each dimension of a point in D is normalized to $[0,1]^d$ .

### A. K-means Algorithm in MapReduce

MapReduce is a parallel programming framework for large-scale data sets that abstracts parallel computing processes into two functions: Map and Reduce. Both the Map function and the Reduce function take a <key, value> pair as input, and after processing, convert it to another batch of <key, value> as output. In the MapReduce Framework, the data set is split into many small data blocks which are first passed to the Map function for processing, then its result is used as the input of the Reduce function.

The core of the K-means clustering algorithm is to determine the cluster to which a data point belongs, by calculating the distance between that point and the centroids. In MapReduce, firstly use the Map function to calculate the distance between each point and the centroids, and select the centroid with the smallest distance as its cluster. Each point is represented in the format of <key, value> as <the cluster center identifier to which the point belongs, and the attribute vector of the point>. The Reduce function will receive points with the same key value, namely points belonging to the same cluster, and calculate new centroids based on these points. The specific calculation process of the K-means algorithm in the MapReduce framework is shown in **Algorithm 1**.

**Algorithm 1: PK-Means Clustering**

**Input:** dataset D, number of clusters *k* and iteration stop condition

**Output:** k clusters

(1) Randomly select *k* points in the dataset D as the initial centroid;

(2) Divide the dataset D into *n* disjoint subsets $D_1, D_2, \cdots, D_n$ in n Map tasks;

(3) In each Map task $Map_i$, calculate the distance between each point in $D_i$ and the centroids, then select the centroid with the smallest distance as its cluster;

(4) Each Map task $Map_i$ produces a list of <key, value> pairs from $D_i$ which is sent to the Reduce function.

(5) Each Reduce task accepts <key, value> pairs with the same key value and calculates a new centroid $o^i$ .

(6) Determine whether the iteration continues by testing the stop condition. If continue, repeat steps (3)(4)(5); otherwise, go to (7);

(7) Stop the clustering process and output the clustering result.

### B. Adding Noise

In order to achieve the differential privacy, so that the points in the data set change without revealing the information of a single point, we use Laplace mechanism to add noises in the process of calculating the centroid. The Laplace mechanism compute a noisy result of function $f$ on the dataset D by adding to $f(D)$ a random noise, as shown in the following equation:

$$A_f(D) = f(D) + Lap(\frac{GS_f}{\varepsilon}), (2)$$

Where

$$\Pr[Lap(\beta) = x] = \frac{1}{2\beta} e^{-|x|/\beta} .$$

And $GS_f$ is the global sensitivity of function $f$ , given as the following equation:

$$GS_f = \max \| f(D) - f(D') \|,$$

where D and D′ differ in at most one data record.

The Reduce task first calculates the number of points included in task *C* and the sum of the coordinates of the data points in the i'th dimension $S_i$ . The noise is added to *C* and $S_i$ respectively, and get the obfuscated $C'$ and $S_i'$ . Then, a new centroid is obtained by $o^j = S_i' / C'$ .

Another important issue is the allocation of privacy budgets. The choice of the number of iterations directly affects the allocation of the privacy budget. There are generally two ways to determine the number of iterations, which correspond to two different methods for the allocation of privacy budgets in each iteration of the clustering algorithm. One way is to fix the number of iterations. In some literature such as [44], the number of iterations is artificially determined with equal privacy budget allocation for each round. Another way is proposed in [45], in which the number of iterations is uncertain, each iteration consumes half of the remaining privacy budget. Considering that as the iterations proceed, the harm to the accuracy of results will increase with the privacy budget decreasing, we adopt the former method in which the number of iterations is fixed and the privacy budget to each iteration is equally allocated to improve the accuracy of the clustering results.

Based on the two ways of determining the number of iterations as described above, there are two main methods for the allocation of privacy budget. One is that the number of iterations is uncertain and the privacy budget for iteration t is $\varepsilon/2^{t+1}$ ; the other method is first to determine the number of iterations *T*, then the privacy budget for each iteration is $\varepsilon/T$ .The first method can greatly reduce the accuracy of the result as the round number increases and the privacy budget decreases. Therefore, in terms of budget allocation, we choose the second method. The number of iterations is determined by the method introduced in [34], which considers MSE between the noisy centroids and the true centroids in one iteration. The MSE of a noisy centroid $\hat{o}$ is calculated as:

$$MSE(\hat{o}) = E\left[\sum_{i=1}^{d}\left(\frac{S_i+\Delta S_i}{C+\Delta C}-\frac{S_i}{C}\right)^2\right].$$

Where $C$ is the number of data points in the cluster, $S_i$ is the sum of the coordinates of data points in the i'th dimension, $\Delta C$ is the noise added to $C$ and $\Delta S_i$ is the noise added to $S_i$. On the i-th dimension,

$$MSE(\hat{o}_i) = E\left[\left(\frac{S_i+\Delta S_i}{C+\Delta C}-\frac{S_i}{C}\right)^2\right]$$
$$= \frac{Var(\Delta S_i)}{C^2}+\frac{S_i^2 Var(\Delta C)}{C^4}.$$

In this paper, we set the range of each dimension to be [0,1]. We suppose that on average $\rho = S_i/C$ and $C \approx N/K$. Hence, $MSE(\hat{o}_i)$ can be approximated as follows:

$$MSE(\hat{o}_i) \approx \frac{k^2}{N^2}\left(Var(\Delta S_i)+\rho^2 Var(\Delta C)\right). (3)$$

As the Laplace noise added to each dimension is independent, from Equation 3 we know

$$MSE(\hat{o}) \approx \sum_{i=1}^{d} MSE(\hat{o}_i)$$
$$= \frac{k^2 d}{N^2}\left(Var(\Delta S_i)+\rho^2 Var(\Delta C)\right)$$
$$= \frac{k^2 d}{N^2}\left(\frac{2}{\varepsilon_i^2}+\rho^2\frac{2}{\varepsilon_0^2}\right).$$

Where $\varepsilon_i$ is the privacy budget allocated to the i-th dimension, $\varepsilon_0$ is the privacy budget allocated to $C$ and $\sum_{i=1}^{d}\varepsilon_i+\varepsilon_0=\varepsilon_t$. The global sensitivity of $S_i$ and $C$ are both 1. Therefore, $Var(\Delta S_i)$ is $2/\varepsilon_i^2$ and $Var(\Delta C)$ is $2/\varepsilon_0^2$. Because the range of each dimension is [0,1], we can get $\varepsilon_i : \varepsilon_0 = 1:1$ and

$$\varepsilon_i = \varepsilon_0 = \frac{\varepsilon_t}{d+1}$$

Then we can obtain the MSE of all the centroids

$$MSE(\hat{O}) = \sum_{j=1}^{k} MSE(\hat{o}^j)$$
$$= \frac{k^3 d}{N^2}\left(\frac{2}{\varepsilon_i^2}+\rho^2\frac{2}{\varepsilon_0^2}\right)$$
$$= 2(1+\rho^2)\frac{k^3 d}{N^2\varepsilon_0^2}$$
$$= 2(1+\rho^2)\frac{k^3 d(1+d)^2}{N^2\varepsilon_t^2}.$$

Let the sum of MSE of all the centroids be no more than 0.01, we can get the minimal privacy budget $\varepsilon_m$ allocated to one iteration. It follows the Equation 4 that

$$2(1+\rho^2)\frac{k^3 d(1+d)^2}{N^2\varepsilon_t^2} \le 0.01. (4)$$

We can find that

$$\varepsilon^m = \left(\frac{200k^3 d(1+d)^2}{N^2}\left(1+\rho^2\right)\right)^{\frac{1}{2}}.$$

Then we can determine the number of iterations based on $\varepsilon_m$. We use the method proposed in [34]. For $\varepsilon \le 2\varepsilon_m$, we set the number of iterations $T$ to be 2, and the privacy budget allocated to each iteration is $\varepsilon/2$. For $\varepsilon > 2\varepsilon_m$, $T$ is determined by the following equation:

$$T = \min\left\{7, \frac{\varepsilon}{\varepsilon_m}\right\}. (4)$$

The privacy budget allocated to each iteration is $\varepsilon / T$.

### *C. Selecting the Initial Centroids*

An important issue in the k-means clustering algorithm is the choice of the initial centroids. The clustering results will fluctuate with the initial centroids. In the traditional k-means algorithm, the initial centroids are randomly selected in the data set. Starting with different initial cluster centers will result in different clustering results, and it is easy to fall into the local optimal solution. Selecting a suitable set of initial centroids can be very helpful in improving the accuracy and stability of the results.

In this paper, we use an improved canopy algorithm, **Algorithm 2**, to select the initial centroids for K-means algorithm and implement it in the MapReduce framework. We set two thresholds $t_1$ (the loose distance) and $t_2$ (the tight distance) for the canopy algorithm. Considering that the calculation time of the canopy algorithm is too long when the volume of the data is large, we first randomly select a subset $D_C$ of the dataset D as the input of the canopy algorithm. We set $|D_c| = 20k$. Then, in the MapReduce, we use n Map tasks to decide the canopy to which each data point in $D_C$ belongs and use a Reduce task to compute the final canopies. In this way, we obtain the canopies and the data point set in each canopy. Since the number of canopies may be more than the number of clusters in the k-means algorithm, we select the first k canopies $\{C_1, C_2, \cdots, C_k\}$ with the largest number of data points to calculate k initial centroids and add noise to the centroids using the following equation:

$$\hat{o}_i^j = \frac{\sum_{x^m \in C_j^{t_2}} x_i^m + Lap(\frac{1}{\varepsilon_i})}{\left|C_j^{t_2}\right| + Lap(\frac{1}{\varepsilon_0})}. (5)$$

Where $i = 1,2,\cdots,d$, $j = 1,2,\cdots,k$, $C_j^{t_2}$ stands for a set of points in the j-th canopy that the distance from the center of j-th canopy is less than $t_2$. And $\bar{o}_i^j$ is the i-th dimension of the j-th centroid.

We take the selection of initial centroids as the first iteration. So, in this process, the total privacy budget equals to $\varepsilon/T$, which is the same as the other iterations in K-means process. Additionally, considering that this process does not involve the same points since we use $C_j^{t_2}$ to calculate the centroids, the noise added in equation 4 is the same as the noise added in K-means iterations.

**Algorithm 2: Selecting Initial Centroids**

**Input:** dataset D, the number of centroids *k*, two thresholds $t_1$ (the loose distance) and $t_2$ (the tight distance) where $t_1 > t_2$

**Output:** *k* centroids

1.Begin with the dataset D to be clustered.

2.Randomly select a subset $D_C$ of D as the input.

3.Remove a point from $D_C$ and begin a new 'canopy'.

4.For each point left in $D_C$, assign it to the new canopy if the distance is less than $t_1$.

5.If the distance of the point is also less than the tight distance $t_2$, remove it from the original set $D_C$.

6.Repeat step 2-5 until there are no more data points in the set $D_C$.

7.Select the first *k* canopies $\{C_1, C_2, \cdots, C_k\}$ with the largest number of points.

8.Calculate the centroids and add noise.

9.Output the *k* centroids.

## VI. Privacy Analysis

Differential privacy has two characteristics: sequence combination and parallel combination, both play an important role in the allocation of privacy budget. If there are m random algorithms $A_1, A_2, \cdots, A_m$, and $A_i$ (1 ≤ i ≤ m) satisfies $\varepsilon_i$-differential privacy, then for the same data set D, the sequence combination algorithm $\{A_1, A_2, \cdots, A_m\}$ also satisfies $\varepsilon$-differential privacy, in which $\varepsilon = \sum_{i=1}^{m} \varepsilon_i$. If there is a random algorithm $M$ and a dataset D, in which D is divided into disjoint subsets $D_1, D_2, \cdots, D_n$. If algorithm M satisfies ε-differential privacy, then the algorithm composed of the combination operation of M on $\{D_1, D_2, \cdots, D_n\}$ also satisfies $\varepsilon$-differential privacy.

As described in section III, the privacy of the k-means algorithm that satisfies differential privacy in the MapReduce framework is achieved by adding Laplace noise to the results of count and sum operations in each Reduce task. Since the process of selecting initial centroids and each iteration of k-means algorithm are equivalent to the sequence combination of the random algorithm, the privacy budget of the entire algorithm is

$$\varepsilon = \sum_{t=1}^{T} \varepsilon_t .$$

Where *T* is the total number of iterations and $\varepsilon_t$ is the privacy budget for the *t*-th iteration. The privacy budget for each iteration is $\varepsilon/T$. In each iteration of K-means algorithm, since Reduce tasks are performed independently, the results of each iteration are equivalent to the parallel combination of the Reduce tasks. Therefore, in order to satisfy the $\varepsilon_t$-differential privacy in the *t*-th iteration, it is necessary to make each Reduce task in the distributed environment satisfy $\varepsilon_t$-differential privacy. That is to say, in one iteration, the privacy budget of each Reduce task is the same.

In the Reduce task of one iteration, *d+1* noise will be added, including **C** and $S_i$ of d dimensions. Since one point is added or deleted to the data set, the maximum change of *C* is 1, the global sensitivity of count query is $GS_C = 1$. If dataset D is normalized to [0, 1], when adding or deleting a point from the dataset, the maximum change of each attribute $S_i$ is 1. The global sensitivity of $S_i$ is $GS_{S_i}$. According to equation 2, in each iteration, adding noise $Lap(1/\varepsilon_0)$ to *C* and adding noise $Lap(1/\varepsilon_1)$ to $S_i$ can make the algorithm satisfy differential privacy, where $\sum_{i=1}^{d} \varepsilon_i + \varepsilon_0 = \varepsilon_t$. Considering that when the range of each dimension is [0,1], the privacy budget allocated to $S_i$ and *C* satisfy $\varepsilon_i : \varepsilon_0 = 1:1$. So

$$\varepsilon_i = \varepsilon_0 = \frac{\varepsilon_t}{d+1},$$

and the noise added to *C* and $S_i$ is $Lap(\frac{1}{\varepsilon_t/(d+1)})$.

## VII. Performance Evaluation

The main function of the scheme proposed in this paper is to use Laplace mechanism and an improved initial point selecting method to preserve data privacy and increase the accuracy of the algorithm. The MapReduce distributed computing framework is used to improve the efficiency of K-means clustering algorithm. The privacy of the algorithm has been demonstrated in section 6. In this section, we conduct experiments to measure NICV and the efficiency of the algorithm for performance evaluation.

TABLE I. Description of Dataset

| Dataset | *tuples* | *dimension* | *clusters* |
|---|---|---|---|
| **Blood** | 748 | 4 | 2 |
| **Adult** | 48842 | 6 | 5 |

We experimented with two datasets Blood and Adult from the UCI Knowledge Discovery Archive database. The dataset Blood contains individual information of blood donation and the dataset Adult contains the identity of the individuals and other general information. Table 1 summarizes the two datasets. For the dataset Blood, the number of records is 748, the dimension of records is 4 and the number of clusters is 2. We set k=2 for this dataset. And we create a new attribute representing the classification result to evaluate the accuracy of our algorithm. For the dataset Adult, the number of records is 48842. We choose 6 continuous variables as the attributes in records. We set k=5 for this dataset according to variable "race" in the original dataset. And we set $\rho$=0.225 [34].

In the experiment, the cloud computing platform consists of a computer as the master node and two other computers as the worker nodes. The Hadoop is deployed into our cloud computing platform. The experimental environment with one master node and two node nodes is set up as follows. CPU: Intel Core i7-6700 3.40GHz; RAM: 8GB; System: Linux. The clustering algorithm is developed in Java.

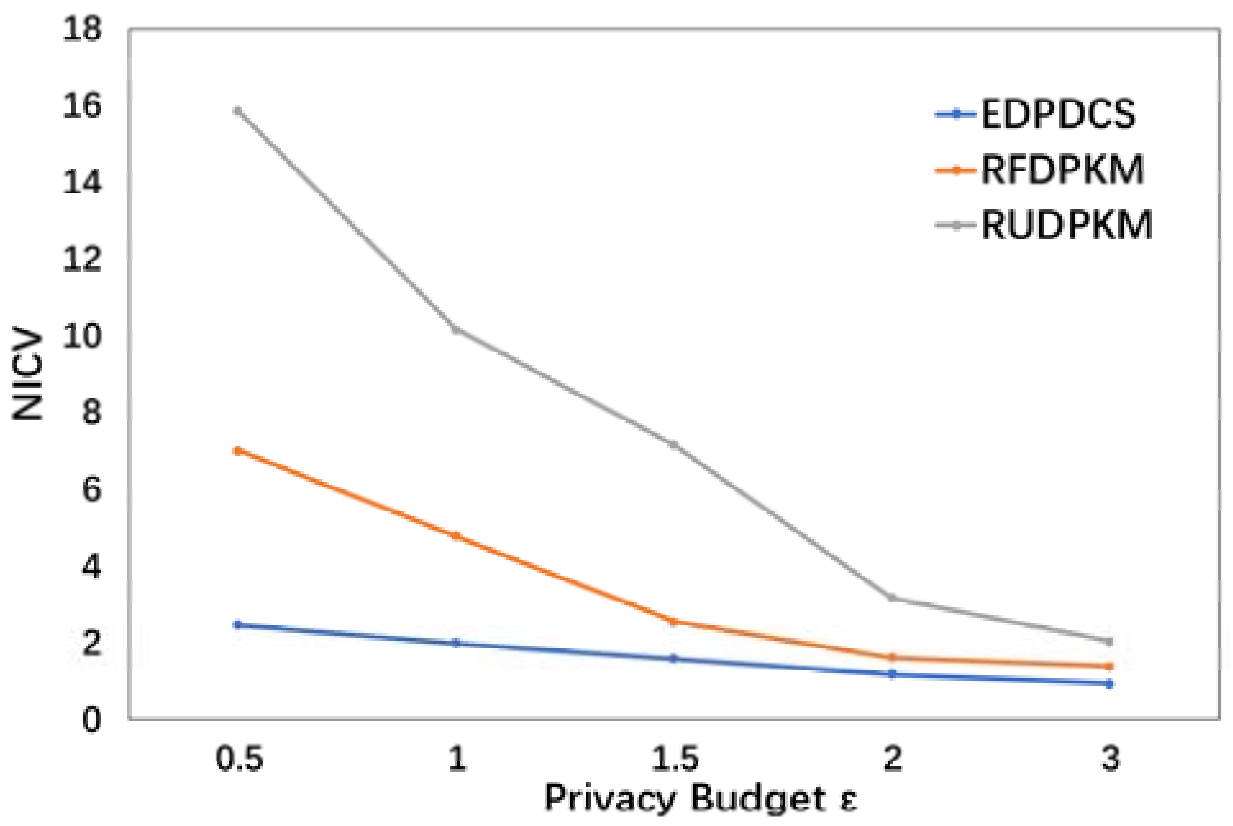

Fig. 3. NICV of our proposed EDPDCS, RFDPKM and RUDPKM on dataset Blood with different privacy budget

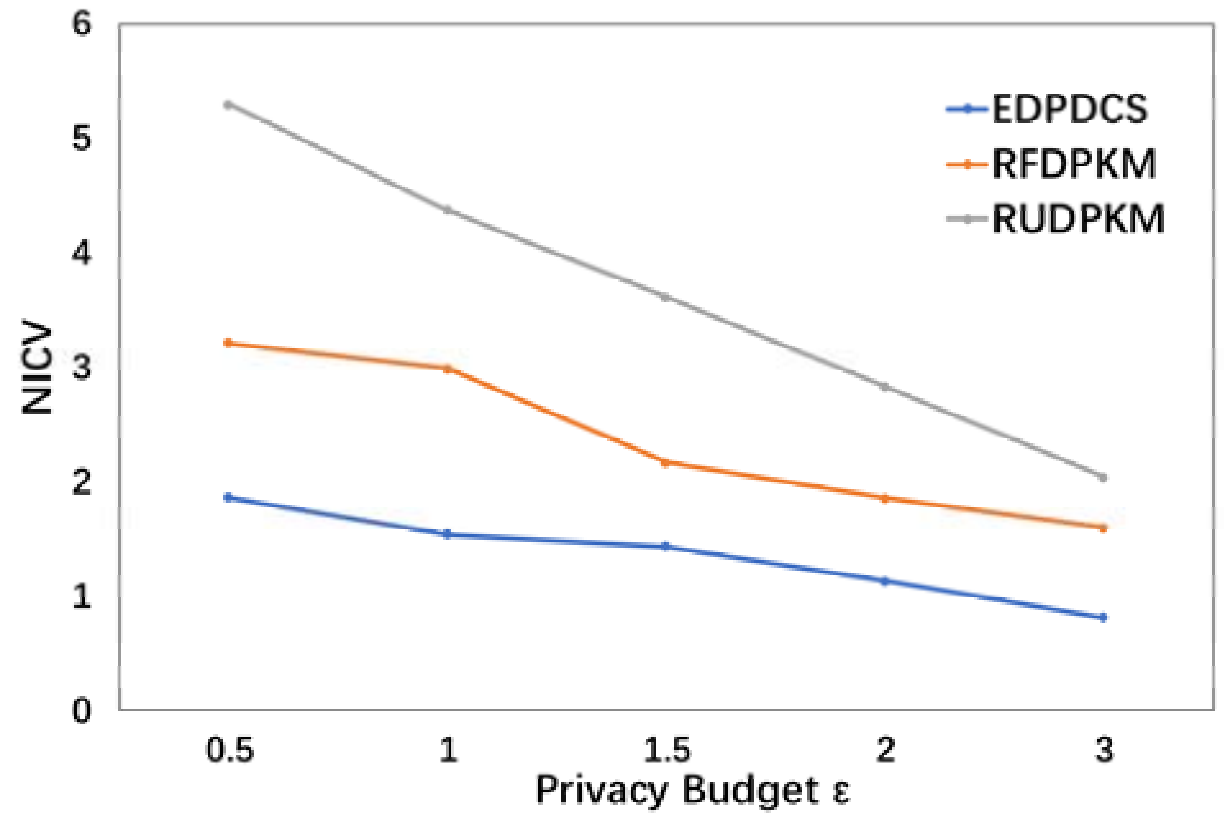

Fig. 4. NICV of our proposed EDPDCS, RFDPKM and RUDPKM on dataset Adult with different privacy budget

## A. Accuracy

In this subsection, we compare the NICV of three clustering algorithms through experimental testing. The research direction of improving the accuracy of clustering algorithms satisfying differential privacy mainly focuses on two aspects. The first is to improve the selection method of the initial centroids. The other is to optimize the allocation of privacy budgets in the iterative process of the algorithm. We compare our proposed algorithms in these two aspects with other two algorithms. The first algorithm is called RU DP K-means, in which the initial point is selected randomly, and the number of iterations is uncertain. The second algorithm is called RF DP K-means, in which the initial point is selected randomly, and the number of iterations is fixed.

TABLE II. The Number of Iterations

| | $\varepsilon=0.5$ | $\varepsilon=1$ | $\varepsilon=1.5$ | $\varepsilon=2$ | $\varepsilon=3$ |
|---|---|---|---|---|---|
| **Blood** | 2 | 2 | 2 | 3 | 4 |
| **Adult** | 7 | 7 | 7 | 7 | 7 |

According to the algorithm proposed above, we can calculate the minimal privacy budget allocated for each iteration. For the dataset Blood, $\varepsilon_m$=0.65508. And for the dataset Adult, $\varepsilon_m$=0.06799. In order to observe the impact of privacy budget on the availability of clustering result, we test with several different total privacy budget 0.5, 1, 1.5, 2, 3. The number of iterations determined by equation 4 for datasets Blood and Adult are shown in Table II. And results are shown in Fig.3 and Fig.4. It can be seen from the experimental results that the NICV values of our algorithm are smaller than the other two algorithms, proving that our algorithm outperforms the other two clustering algorithms. In addition, when the privacy budget is small, that is, the degree of privacy preservation is strong, the accuracy of the clustering results of our scheme is significantly higher than the other two algorithms.

## B. Efficiency

In this subsection, we compare the efficiency of the algorithms implemented in MapReduce. In order to evaluate the impact of the number of distributed computing nodes to the efficiency of the algorithm, we conduct experiments with 1 and 3 nodes respectively, and measure the running time of our proposed algorithm under different size of data points.

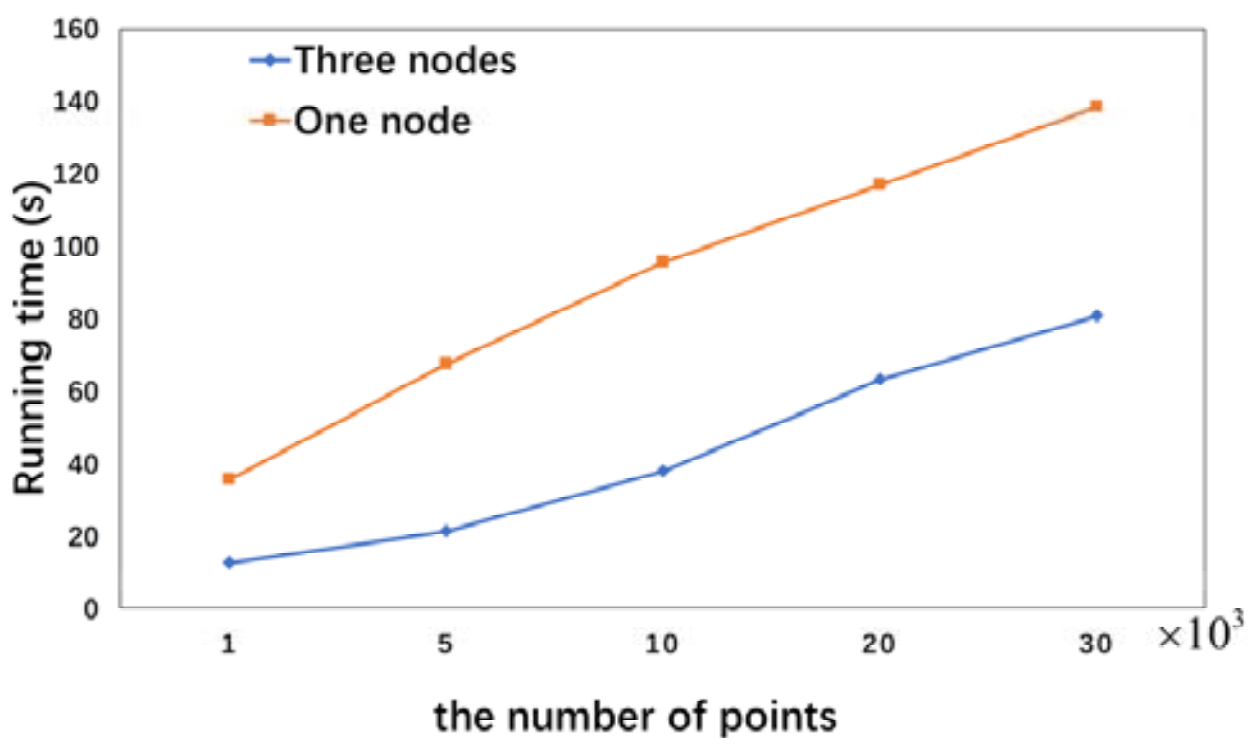

Fig.5. Comparison of running time with different number of nodes and different size of data volume

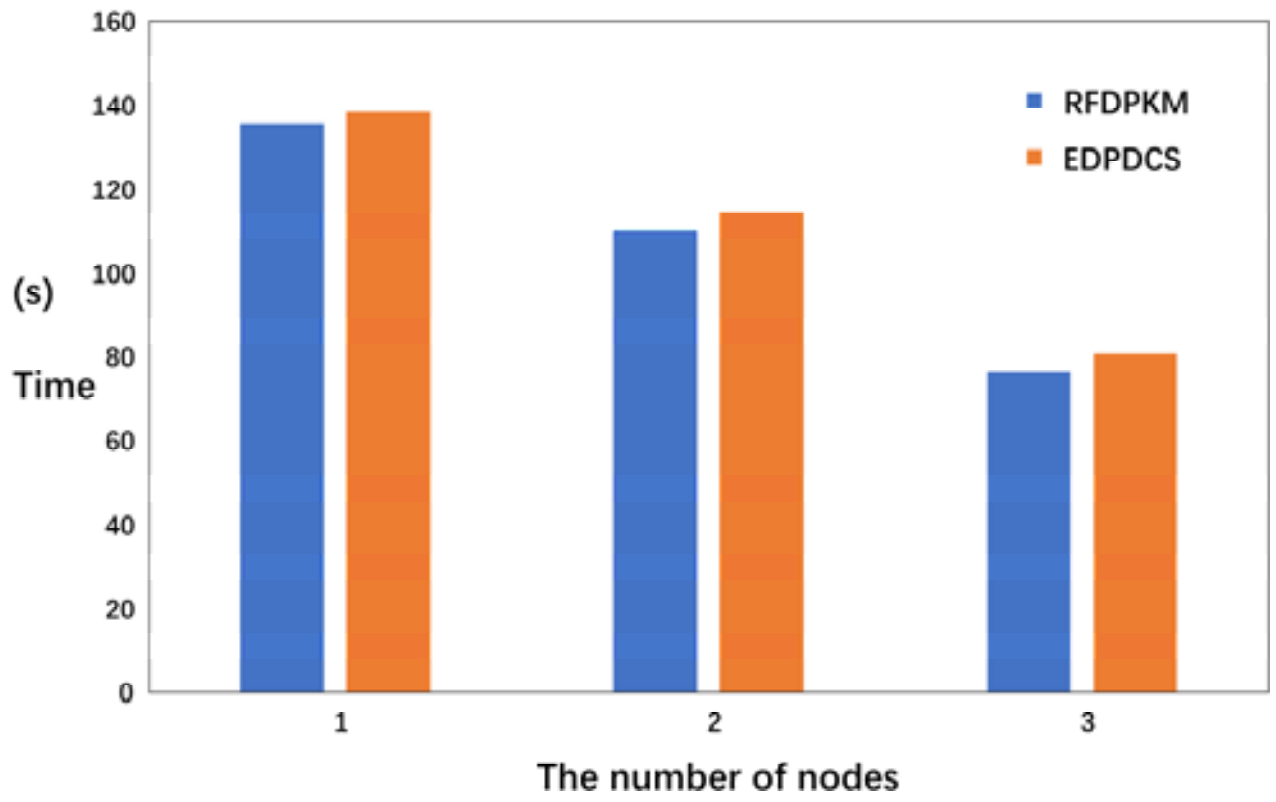

Fig. 6. Comparison of running time using REDPKM and EDPDCS on dataset Adult with different number of nodes

This can also evaluate the influence of the data volume on the efficiency. Additionally, we conduct another set of experiments to compare the running time of our proposed algorithm with RFDPKM, with 1 up to 3 nodes. The experimental results are shown in Fig.3 and Fig.4 The running time of our algorithm decreases when more nodes are used. That is to say, the MapReduce framework has improved the efficiency of our algorithm. Besides, from the second set of experiments, we demonstrate that our proposed algorithm is as efficient as RFDPKM.

## VIII. Conclusion

In this paper, we propose an efficient differentially private data clustering scheme for IoMT. In the proposed scheme, we optimize the allocation of privacy budgets and the selection of initial centroids to improve the accuracy of differentially private K-means clustering algorithm. Specifically, the number of iterations of the K-means algorithm is set to a fixed value determined by the total privacy budget and the minimal privacy budget allocated to each iteration calculated by analyzing the mean squared error (MSE) between noisy centroids and true centroids in one iteration. In addition, an improved initial centroids selection method, selecting a small portion of the dataset and performing rough clustering in advance to select the initial centroids, is proposed to increase the accuracy and efficiency of the clustering algorithm. We deploy this algorithm in the MapReduce framework. Experiments are conducted to compare the NICV in the clustering results of our method with another two methods. The evaluation results show that our proposed algorithm can improve the accuracy of the differentially private k-means algorithm while preserving privacy for data contributors in IoMT.

## Acknowledgment

This work is supported by Beijing Natural Science Foundation under grant 4182060.